# High-Temperature Superconductivity Mechanism for Cuprates

Richard H. Squire [κ], Norman H. March [*,ξ,†]

[κ] Department of Chemistry, West Virginia University
Institute of Technology
Montgomery, WV 25303, USA

[*] Department of Physics, University of Antwerp, Groenborgerlaan 171,
B-2020 Antwerp, Belgium

[ξ] Oxford University, Oxford, England

[†] Abdus Salam International Center for Theoretical Physics
Trieste, Italy

ABSTRACT

Egorov and March's [1] plotted the product of resistivity and the copper spin-lattice relaxation $T_1$ vs. temperature for $YBa_2Cu_4O_8$ finding a minimum at temperature $T_m > T_c$, the superconducting temperature, heralding an electronic phase change which can be interpreted as the formation of a "preformed" pair. In this context we propose a superconducting mechanism based on the notion that the preformed pair is a "soft" boson (a localized, different type of Cooper pair) which dissociates above the classical BKT transition temperature, resulting in two circular charge density waves. The model suggests explanations for considerable experimental work and offers a physical explanation for the basis of the Uemura-Homes law.

## 1. Background.

Early work by one of us [2] emphasized the importance for the properties of the hole liquids flowing through initial antiferromagnetic assemblies of Cu spins in the normal state of high $T_c$ superconductors (SC). This proposal has been strongly supported by the recent experimental study of Jin et. al. [3]. Subsequently, Egorov and March [1] used normal state experimental values of electrical resistivities R and nuclear spin-lattice relaxation time $T_1$ at Cu sites in $YBa_2Cu_4O_6$ to plot the product $RT$ vs. temperature $T$. A linear relationship between $RT$ and $T$ is in evidence over a temperature range from $150-450K$ for this superconductor (see Fig 2 in [4b]). As the sample was further cooled, there was a minimum around $100K$ as the superconducting transition temperature $T_c$ was more closely approached, subsequently interpreted as due to an electronic phase change resulting in preformed CPs (bound BKT vortices) above $T_c$. This possibility had been predicted in the theoretical work of Nozieres and Schmitt-Rink [5] prior to the discovery of high-$T_c$ superconductors. In their review on spin fluctuations and high $T_c$ superconductivity, Motiya and Ueda [6] write that the relaxation rate $1/T_1$ associated with Cu sites is quite different from the conventional Korringa relaxation for metals, namely $1/T_1T = const$. They conclude that this relaxation rate for high-$T_c$ cuprates saturates at elevated temperatures so that $1/T_1T$ behaves similarly to the Curie-Weiss law, a conclusion was already anticipated by Egorov and March [3].

It has been speculated by some [7] that details regarding the nature of the preformed CP in high-$T_c$ models may not be necessary since the coherence length is comparatively small and therefore takes place in coordinate space. Indeed, progress has resulted by use of a very general theory based on a "hard core" boson field $\phi(r)$, where the internal structure is completely ignored. A number of features can be understood with this approach, but a more complete explanation of important features such as the itinerant (delocalized) Cooper pairing mechanism, the pseudogap and the Nernst effect require a detailed structure of the preformed Cooper pair. The purpose of this manuscript is to describe the relevance of the "soft" boson preformed pair which could have a more general occurrence.

The outline is as follows; first we summarize some experimental and theoretical boundaries a successful cuprate model must have in Section 2 along with our proposed "soft" boson model. Section 3 is a brief discussion of the BKT theory, the model basis, while section 4 offers an explanation of the SC dome and the pseudogap region of the phase diagram including the 2D to 3D transition. Comparison with other similar phenomena is contained Section 5, followed by a summary, conclusions, and proposals for future study in Section 6. An Appendix discussing "soft" bosons follows.

## 2. Experimental and Theoretical Boundaries for Cuprate Model.

**2.1. Experiment.** Jin et al [2] presents evidence relating an electron liquid flowing through an antiferromagnetic assembly discussed earlier. More specifically, in studies of electron-doped cuprate SC's (no pseudogap), there is a linear dependence of the scattering rate on temperature that correlates with electron pairing above $T_c$ and it continues to zero temperature. Additional confirmation of local electron pairing is also evident in the STM data of Kohsaka et al [8], providing data on cuprate SC in very low doping conditions. Using quasi-particle interference (QPI) imaging that allows them to determining the material's electronic structure in k- and r- space simultaneously, they observe the transfer of spectral weight to the higher energy r-space states which are localized Cooper pairs, breaking translational and rotational symmetries at the atomic scale. Delocalized Cooper pairs vanish as the Mott insulator is approached by removing a few holes, leaving "localized" Cooper pairs. Additional STM studies by Pan et al [9] suggest vortices in other HTSC materials (YBCO and BSCCO) as a common phenomenology. Lastly, a model needs to satisfy the empirical Uemura-Holmes laws [10, 11] which states that $T_c$ is proportional to the superfluid density $\rho_s$ near $T \square 0$ multiplied by the electrical resistivity $\rho_{dc}$ just above $T_c$ (see Section 4 for discussion).

**2.2. Theory.** To focus only on crucial correlation terms in as rigorous a Hamiltonian operator as practical, we use an electron density form as formulated by Bohm and Pines [12], and Turlakov and Leggett [13], namely

$$\hat{H} = \sum_{p,\sigma} \frac{p^2}{2m} c^\dagger_{p,\sigma} c_{p,\sigma} + \frac{1}{2\Omega} \sum_q V_q \left[ \hat{\rho}_q \hat{\rho}_{-q} - N \right] + \sum_{\kappa \neq 0} U_{-\kappa} \hat{\rho}_\kappa \qquad (2.1)$$

with

$$\hat{\rho}_q = \sum_{k,\sigma} c^\dagger_{k-q,\sigma} c_{k,\sigma} = \sum_{\vec{r}_i} e^{-i\vec{q}\vec{r}_i} \qquad (2.2)$$

The $c^\dagger_{k,\sigma} c_{k,\sigma}$ are the creation and annihilation operators, respectively, the total density operator is $N$, the number of electrons, $\Omega$, the volume, and $\rho_{\vec{k}}$ is the Fourier transform of the density operator. The first and second terms are the kinetic and coulomb terms of only the electrons; the last term is a lattice interaction where $U_{-\kappa}$ is an Umklapp potential interacting with electron density summed over reciprocal lattice wave vectors $\kappa$. Within the limits of the Hamiltonian above, there are at least two theoretical conditions which any proposed solution of the cuprate HTSC problem needs to address.

1) Turlakov et al have used the above Hamiltonian to establish sum moment rules which restrict the upper and lower bounds on electron Coulomb energy, thereby restricting the possible models for the cuprate superconductors. One possibility meeting this requirement is a band in the mid-infrared (MIR) region of the spectrum; high $T_c$ cuprates satisfy this requirement by showing absorption from 40 to 200 meV (MIR region). Wulin et al found this broad peak to be highly correlated with their interpretation that the pseudogap results from preformed pairs. They conclude that their model is consistent with the sum rules [14].

2) If a two dimensional solution is proposed, it needs to be local and then perhaps expand into the third dimension for superconductivity to satisfy the Mermin-Wagner-Hohenberg-Coleman (MWHC) theorem [15] which states that long range order is not possible in one or two dimensions.

**2.3 Proposed Model for a Preformed Cooper Pair.**

It is well known that instability near the Fermi energy caused by an attraction affects both the lattice and the electron gas. This attraction can be a fluctuation of the electron density that under certain conditions undergoes a phase transition to a charge density wave (CDW) [16]. The result is a frozen lattice distortion in a macroscopically occupied phonon mode that can be described by an order parameter

$$|\Delta|e^{i\phi} = g\left(\langle b_{2k_F}\rangle + \langle b^{\dagger}_{-2k_F}\rangle\right) \tag{2.3}$$

The periodic boundary conditions can represent a circular system and the same equations obtain as in a typical linear CDW solution. The circular solution (circDW) can be viewed as a very weak pole in the density-density correlation function eq (2.1) and have some features in common with rotons (see Section 5.1). After substitution in the Fröhlich Hamiltonian, some approximations, and diagonalization by a Bogoliubov-Valatin transformation, the resulting binding energy of a circDW has the same form as a superconductor [16],

$$\Delta = 2\varepsilon_F e^{-1/\lambda} \qquad \lambda = \frac{g^2 n(\varepsilon_F)}{\hbar \omega_{2k_F}} = g'n(\varepsilon_F) \tag{2.4}$$

where $\lambda$ is a dimensionless electron-phonon coupling constant resulting in a correlation energy of approximately

$$E_{cond} \simeq E_{norm} - E_{CDW} = \frac{n(\varepsilon_F)}{2}\Delta^2 \tag{2.5}$$

A material containing circDW's could be an insulator if two circDW's are bound; on the other hand if they are not, they might be a significant factor in conductivity as other have suggested previously [17]. Based on the STM evidence above, we view the circCDW as a vortex and propose the Egorov-March phase change follows as the temperature is lowered where two vortices become paired following a modified BKT theory (Section 3). This is a different notion of a Cooper pair with a "d" symmetry structure (see Section 5 for analogies) and a possibly larger energy/smaller size can result (see Appendix A). As mentioned, preformed pairs were predicted by Nozieres and Schmitt-Rink [5], and augmented by Ohashi and Griffin [7c] to include two types of Cooper pairs, one pair localized and the other mobile. As the charge of the proposed vortex pair is now 2e, we call this vortex pair a Circ2DW or a localized Cooper pair (LCP). Being boson-like in nature, the LCP/Circ2DW will suppress the single particle spectrum as observed in the cuprates.

## 3. Vortices.

### 3.1. XY Model and the Berenzinskii-Kosterlitz-Thouless (BKT) transition [17, 18, 19, 20, 21, 22].

The novel work resulting in the BKT Theory describes either classically interacting spins on a 2D lattice or a 2D lattice gas model of a superfluid with similar results. At low temperatures an order parameter $\psi = \rho e^{i\theta(\vec{r})}$ described by a magnitude $\rho$ and a phase $\theta_i$ is associated with each lattice site i and small fluctuations of the magnitude and phase can be written as

$$\frac{H}{k_B T} = -J \sum_{\langle i,j \rangle} \cos(\theta_i - \theta_j) \qquad (3.1)$$

where J is an energy scale. Expanding to second order (Gaussian approximation)

$$\frac{H}{k_B T} = const. + \frac{J}{2} \sum_{\langle i,j \rangle} \cos(\theta_i - \theta_j)^2 \qquad (3.2)$$

Jose and others [23] using the Fourier transforms of the angles $\theta_i$, proportional to $\int d^2k \, c(\vec{k}) \exp(i\vec{k} \cdot \vec{R}_j)$, found eq (3.2) is then diagonal in $c(\vec{k})$, and arrived at the result that the spin correlation function for a square lattice is

$$\langle \cos\theta_i \cos\theta_j \rangle = \frac{1}{2} \exp\{-\frac{a^2}{4\pi^2} \iint d^2q \frac{1-\cos(\vec{q} \cdot \vec{R}_{ij})}{4 - 2\cos(q_x a) - 2\cos(q_y a)}\} \qquad (3.3)$$

with $\vec{R}_{ij}$, the vector between i and j. For large values of $R_{ij}$ it can be shown that the integrals depends logarithmically on the ratio $a/R_{ij}$ as

$$\left\langle \cos\theta_i \cos\theta_j \sim \left(\frac{a}{R_{ij}}\right)^{\eta(T)} \right\rangle \qquad (3.4)$$

and

$$\eta(T) = \frac{1}{2\pi J} \qquad (3.5)$$

For a superfluid assuming a pinned normal component, the energy associated with a fluctuation of the superfluid velocity $\vec{v}_s(\vec{r})$ for $^4He$ films is

$$H = \frac{1}{2} \rho_s \iint d^2r \, |\vec{v}_s(\vec{r})|^2 \qquad (3.6)$$

with $\rho_s$ being the aerial (2D) superfluid density which sets the energy scale. Galilean invariance requires

$$\vec{v}_s(\vec{r}) = \frac{\hbar^2 \nabla \theta(\vec{r})}{2m} \qquad (3.7)$$

The energy associated with phase variation (the so-called phase stiffness) in a superfluid is

$$H = \frac{1}{2}\iint d^2r\rho_s \frac{\hbar^2}{m^2}(\nabla\theta)^2 \tag{3.8}$$

where $\theta$ is required to be periodic. Comparing eq (3.1), (3.5) and (3.8), the correlation exponent becomes

$$\eta(T) = \frac{m^2 k_B T}{2\pi\hbar^2 \rho_s}$$

Considering multiple vortices, $\vec{v}_s(\vec{r})$ can be decomposed into

$$\vec{v}_s(\vec{r}) = \frac{\hbar}{m}\nabla\phi(\vec{r}) + \frac{2\pi\hbar}{m}(\hat{z}\times\nabla)\int d^2r' n(\vec{r}')G(\vec{r},\vec{r}') \tag{3.9}$$

with $\phi(\vec{r})$ a nonsingular phase function, $\hat{z}$ a unit vector, $n(\vec{r})$ the vortex charge density and the Green's function satisfying $G(\vec{r},\vec{r}') = \frac{1}{2\pi}\ln\left(\frac{|\vec{r}-\vec{r}'|}{a}\right) + C$ for points far from the boundaries. Substituting eq (3.9) into eq 3.6, we ultimately find two pieces

$$H = \frac{1}{2}J\int d^2\vec{r}(\nabla\phi)^2 + H_{vor} \tag{3.10}$$

The energy is composed of two independent pieces, a topological vortex configuration and assuming that the spin waves are non-singular and independent of the vortices, a spin portion. It can further be shown the BKT theory describes an entity consistent with the MWHC theorem [16]; the algebraic decay of the correlation function suggests that the long range limit must be zero.

### 3.2. Connection with the Free Energy through the Ginsberg-Landau (GL) theory [24, 25]

The "standard" GL course-grained free energy is

$$F = \int d\vec{r}\left(\frac{1}{2}|\nabla\psi|^2 + \frac{1}{2}a|\psi|^2 + b|\psi|^4\right) \tag{3.11}$$

where coefficient a changes signs at a critical temperature and b remains positive. The complex field $\psi(\vec{r})$ is related to the condensate wave function in a superfluid with the probability of a specific configuration being proportional to $\exp(-F/k_B T)$. Rearranging, $|\psi|^2$ is minimized at $|a|/4b$, and setting $|a|/b = const.$ as $b\to\infty$, $\psi(\vec{r})$ can be expressed as

$$\psi(\vec{r}) = \frac{1}{2}\sqrt{|a|/b}\, e^{i\theta(\vec{r})}$$

Substituting into eq (3), the free energy can be expressed with

$$F = const. + \int d^2r\left\{\frac{1}{2}|\nabla\psi|^2 + b\left(|\psi|^2 - \frac{|a|}{4b}\right)^2\right\}$$

Then, the free energy has the form of eq. (3.2) if $J = |a|/(4bk_BT)$. Since vortices do not appear in a power series expansion in T, vortices can be viewed as nonperturbative amplitude fluctuations as the probability of their appearance is proportional to $e^{-E_a/k_BT}$ ($E_a$ is the core energy); the amplitude vanishes at the center and the phase is undefined there also. An isolated vortex in a region L has energy

$$E = \pi k_B T J \ln(L/a) \qquad (3.12)$$

If the possible locations are $(L/a)^2$, the entropy follows

$$S \approx 2k_B \ln(L/a) \qquad (3.13)$$

It can then be shown that the free energy $(F = E - TS)$'s sign at temperature

$$J(T_c \approx 2/\pi)$$

Expressing the coupling $J$ in terms appropriate to a 2D superfluid

$$\frac{\hbar^2 \rho_s^0(T)}{m^2 k_B T} \geq \frac{2}{\pi} \qquad (3.14)$$

which becomes exact using the renormalized superfluid density, $\rho_s^R(T)$.

### 3.3. BKT Renormalization.

**3.3.1. Classical BKT Analysis.** Kosterlitz focused on the correlation which explicitly provides the superfluid density $\rho_s^R(T)$, where R indicates it is renormalized. He found not only a jump discontinuity in the superfluid density [26], but Nelson and Kosterlitz later proved it was universal [27]. Hohenberg and Martin [28] utilized autocorrelations of a superfluid momentum current to define the superfluid density $\rho_s^0(T)$, here unrenormalized by vortex excitations

$$\vec{g}_s = \frac{\delta H}{\delta \vec{v}_s} = \rho_s^0 \vec{v}_s(\vec{r})$$

The correlation function is

$$C_{\alpha\beta}(\vec{q}, K, y) \equiv \left(\frac{\hbar}{mk_BT}\right)^2 \langle \hat{g}_s^\alpha(\vec{q}) \hat{g}_s^\beta(-\vec{q}) \rangle, \qquad (3.15)$$

$\hat{g}_s^\alpha(\vec{q})$ being the $\alpha$ component of the Fourier transform of $\vec{g}_s$ which is the momentum density operator on a microscopic level, $\hat{g}_i(\vec{r}) = \frac{\hbar}{2i}[\hat{\psi}^\dagger(\vec{r})\partial_i\hat{\psi}(\vec{r}) - \hat{\psi}(\vec{r})\partial_i\hat{\psi}^\dagger(\vec{r})]$. To renormalize the superfluid density, the transverse and longitudinal parts of the electromagnetic response are assumed to be separable,

$$C_{\alpha\beta} = A(q)\frac{q_\alpha q_\beta}{q^2} + B(q)\left(\delta_{\alpha\beta} - \frac{q_\alpha q_\beta}{q^2}\right) \qquad (3.16)$$

with $A(q)$ and $B(q)$ depending on parameters K and y. In an isotropic liquid at long wavelengths, $A(q) = B(q)$. As $q \to 0$, the difference between $A(q)$ and $B(q)$ defines the renormalized superfluid density

$$K_R \equiv \frac{\hbar^2 \rho_s^0(T)}{m^2 k_B T} = \lim_{q \to 0}\left[ A(q) - B(q) \right] \quad (3.17)$$

Using eq (3.9) and Fourier transforming and averaging over the phase, $C_{\alpha\beta}$ can be decomposed into

$$A(q) = K \quad (3.18a)$$

$$B(q) = \frac{4\pi^2 K^2}{q^2} \langle \hat{n}(\vec{q}) \hat{n}(-\vec{q}) \rangle \quad (3.18b)$$

where $\hat{n}(\vec{q})$ is the Fourier transform of $n(\vec{q})$. The transverse average in eq (3.18b) is over the vortex portion of eq (3.10). If vortices are bound, the power law decay is rapid, and the order parameter correlation function decays exponentially with the correlation length $\xi_+$,

$$\langle \psi^*(\vec{r}) \psi(0) \rangle \square \exp\left( -r / \xi_+ \right) \quad (3.19)$$

associated with the free vortices density, $n_f$. As mentioned, Nelson and Kosterlitz developed a series of renormalization relationships. One important feature is the correlation length above $T_c$

$$\xi_+(T) \square \exp\left( b' \Big/ |T - T_c|^{1/2} \right) \quad (3.20)$$

which precipitously drops to zero thereby predicting a universal jump discontinuity of the superfluid density irregardless of the film height, substrate, etc. in a number of experiments (Fig 2.8 in [21]). This is impressive when one remembers that the bound vortex pairs are similar to classical molecules [29].

**4. 1. The Transition from 2D to 3D**

The transition to three dimensions of the "soft" boson model generates a rich phase diagram reflecting at least five "zones" of order parameters as the SC dome is traversed with increasing doping. Zone one and three begin and finish with a dominant order, respectively, but zone two (the peak) has interactions which extend into zones four and five in the pseudogap in our model. A qualitative description is first, followed by detail.

As expected, the local "soft" Cooper pairs (LCP's) play a significant role. First, the single-particle tunneling between nearby planes is strongly suppressed by the bosonic nature of the LCP's. Secondly, the characteristic energy scale is set by the phase stiffness at zero temperature as suggested earlier by Emery and Kivelson [30], which is physically expressed in our model by the LCP's (eq 3.8). Lastly, the energy of the vortex seems to control the superconducting $T_c$ [31]. The impact of these statements begins as doping increases along with the superfluid density $n_s$ and $T_c$. Being bosonic in nature, all of the local CP's will be assumed to have the same wave function in their ground state at low temperatures which decays outward from their core in all directions. They are very anisotropic as their x, y dimension is much larger than their z component. Hence, it should be expected that their in-plane properties should reflect this in a "2D SC" (see discussion and figure in [32]). The z component coherent decay from bound vortices will create a finite superfluid density which may assist or interfere with any other coherent components such as Josephson junctions. As the doping and the transition temperature rises, three factors contribute to the peak of the dome: 1) some bosons (LCP's) become excited out of the ground state causing loss of phase coherence, 2) the density of bound vortices will lead to interaction between them, thereby losing their effectiveness and 3) the paired vortices can dissociate into +/- "component" vortices. All of these effects lead to the peak in $T_c$. The dissociation of vortices results in zones four and five where the dynamic qualities of the adjacent portion of the pseudogap where the lower temperature portion of the pseudogap composed of LCP's and CircDW's initially balanced in +/- vortices. The resulting dissipation could result in a net loss of the vortex balance in the electron fluid creating a considerable viscosity. This event will take place over a larger temperature range than the classical BKT universal jump (see below). As the temperature is raised to the higher temperature portion of the pseudogap (zone five), it is suggested that a different phase such as a linear CDW might prevail as certain experiments reveal attributes of a "superconductor" persisting to high temperatures. The peak in $T_c$ has multiple interactions; the predominant contribution to superconductivity in zone 2 is initially a superfluid which competes with resonance interaction as they both exclusively need the LCP to function. As the dome is traversed, the resonance interaction becomes dominant, then collapses into zone 3 where a BCS-like SC prevails. More details follow.

### 4.2.1. Zone 1 of the SC dome.

The density of bound vortices is too low for BEC. We propose fluctuational superfluid patches which control the superconducting transition temperature through "long" range order as a result of overlap with other bosons by phase locking in 3D patches of superfluidity; hence the superfluid density is small in the cuprates. To describe the superfluid density, we first want to examine the contrast with the BCS theory which has a finite gap function $2|\Delta|$ everywhere, so the probability of exciting quasiparticles is quite low if $k_B T \ll 2|\Delta|$, since $n_n(T) \sim e^{-2\Delta/k_B T}$. But in a d-wave cuprate superconductor $\Delta \to 0$ at nodal points on the Fermi surface, so it is always possible to excite quasiparticles with the result that $n_n(T) \sim T$ as experimentally verified [33], on the underdoped side of the dome. Thus, the nodal quasiparticles shown to lead to an initial linear relationship of the superfluid stiffness [34],

$$n_s(T) = n_s(0) - aT \qquad (4.1)$$

This relationship can be compared with a corresponding boson relationship, namely

$$n_s(T) = n_s(0) - 2.612\left(\frac{mk_B T}{2\pi\hbar^2}\right) \qquad (4.2)$$

As is well known, a measurement of the London penetration depth as a function of temperature $\lambda(T)$ can be converted into the superfluid density as a function of temperature $n_s(T)$ using [10]

$$\lambda(T) = \left(\frac{m_e}{\mu_0 n_s(T) e^2}\right)^{1/2} \quad (4.2)$$

As doping increases, the transition temperature $T$ increases until "zone 2" is reached.

### 4.2.2 Zone 2: The boson density remains below the critical BEC level.

The density of superfluidity will drop for the reasons cited in Section 4.1 The transition between zones one and two will be smooth as the boson-fermion Hamiltonian can be rewritten [7a] so that only eigenstates with a macroscopic $N_0$ are relevant for $T < T_c$ and approximate $b_0 \approx N_0^{1/2}$. The vortex pair remains intact and it now becomes a resonant interaction center for spin wave order interacting with the vortex charge order for which we can write an interaction $H_I$ eq (4.3) and add this term to eq (3.10) resulting in eq (4.4),

$$H_I = g_r \sqrt{\frac{N_0}{V}} \sum_{\bar{p},\sigma} [c_{-\bar{p},\sigma} c_{\bar{p}\sigma} \upsilon(p) + h.c.] \quad (4.3)$$

$$\frac{H}{k_B T} = \frac{1}{2} K \int d^2 \vec{r} (\nabla \phi)^2 + \frac{H_{vor}}{k_B T} + H_{int} \quad (4.4)$$

Evidence for the complex comes from three independent sources: first is the work of Mei and Weng [35] who deduce the universal formula eq (4.5) from a spin-roton model of

$$k_B T_c = E_g / \kappa \quad (4.5)$$

classical BKT theory where $E_g$ is the resonance energy in both inelastic neutron and electronic Raman scattering experiments and $\kappa \square 6$. It can be shown that this relationship also applies to our model. Second, we have followed the lead of McDonald and others [36] who found a mapping overlap of spin correlation, Cooper pair correlation, and incommensurability. We added an additional correlation with the checkerboard pattern observed in STM experiments on cuprates and a "d" symmetry for a LCP. The overall result suggests that the interaction center described by eq (4.4) where a spin order interacts with charge order with an offset of $\pi/2$ in the phase as noted by McDonald can account for the charge-spin interaction as Grüner describes (also see fig 6.7, [17]). As doping further increases, this interaction diminishes allowing BCS-like superconductivity to prevail. We discuss the pseudogap next.

### 4.2.3 Superfluid Density Above $T_c$ in the Presence of Disorder.

Recent experiments on the cuprates including Nernst effect [38], finite frequency conductivity [39], and nonlinear magnetization [40], thought to be examples of BKT phase fluctuations, have failed to find the superfluid density universal jump at $T_{BKT}$. Benfatto and others suggest that the downturn is no longer universal, but occurs at a temperature $T_D$ depending on the vortex energy, $E_V$, which seems to depend on the SC model. This contrasts with the universal value for the previously discussed 2D XY model. Their renormalization study suggests that as soon as there is interlayer coupling, the jump is removed and replaced by a downward bend in at $T_D > T_{BKT}$ reflecting the vortex-core energy $\mu$ which scales linearly with $T_c$ [41, 42] (compare with eq 4.5)).

Since it appears from experiment that the vortex core and the superfluid (phase) stiffness, eq 3.8), are interacting, it seems that the previously assumed conserving approximation (gauge invariance) for the electromagnetic kernel $K_{ab}$ breaks down in the classical BKT theory. The BCS approximation is not conserving, but in that theory phase fluctuations only contribute to the longitudinal portion of $K_{ab}$. This is no longer true in the presence of the disorder due to the formation of "soft" boson structures larger than the lattice spacing. The phase couples with both the transverse and longitudinal parts leading to the interaction term above, eq (4.5), as evidence by the $\pi/2$ term in the spin-charge interaction.

The 2D BKT-like transition at temperature $T_D$ combined with Nernst effect experiments on hole-doped cuprates suggests zone four of the pseudogap seems to contain a large region just above $T_c$ where a vortex liquid exists even though the Meissner effect is absent. Some vortices survive above $T_c$ where they dissociate into circCDW's. We can use previous studies of vorticity to encompass the non-ideality present in the non-uniform dissociation. The region is dominated by entropy as previously indicated and seem to have a relatively uniform distribution which can be described by

$$n_{pair} = \left(\frac{1}{a^2}\right) \exp\left[-E_c \Big/ (T-T_c)\right] \qquad (4.6)$$

Experimental evidence suggest the conductance is unusual and contain descriptions such as Orenstein and other's "interpenetrating electrons and vortices" [43], Anguilar and others "pair breaking and coherence peak" in cuprate conductivity [44], Li and others "diamagnetism and Cooper pairing above $T_c$" [45], Anderson's "incompressible vortex" Bose Fluid above $T_c$ [46] and Dubroka's "precursor superconductivity as high as 180K" [47]. It appears that the experimental attributes of a "superconductor" could persist to high temperatures comprising zone 5.

An electrostatic description suggests that below $T_{KT}$, the system has "charge confinement"; it is an insulator. Above $T_{KT}$, bound pairs dissociate and the system is a conductor, becoming a "plasma" of electrons and coherent vortices. In this regard the work of S. Chakravarty and others [48] also merits further studies in the context of the pseudogap phase of the hole-doped cuprates which is proposed to be characterized by a hidden broken symmetry of the $d_{x^2-y^2}-type$. These workers argue that the transition to this state is "rounded by disorder", but in the limit of sufficiently small disorder the pseudogap crossover should reveal itself to be such a transition. It is asserted that the ordered state breaks time-reversal, translational, and rotational symmetries, but remains invariant under the combination of any two. One interesting outcome is the prediction of a metal-metal transition under the superconducting dome. The model presented breaks translational and rotational symmetry in the bound BKT state, and all three symmetries in the proposed pseudogap state.

### 4.2.4 Zone 3: The Fermi liquid returns and the interaction is BCS-like [7a].

With no interaction, there is a degenerate Fermi distribution and no bosons and the highest Fermi energy is $\varepsilon_F = k_F^2/2m < \upsilon$. If $g \neq 0$, $H_1$ continuously extends into a BCS type system with an attraction of initial fermion momenta $\vec{k}$ and $-\vec{k}$ inside the Fermi sea to $\vec{p}$ and $-\vec{p}$ outside and $\upsilon(k) \Box 1 \Box \upsilon(p)$ and the attractive interaction

$$\frac{g^2}{2(\varepsilon_k - \upsilon)} < 0 \qquad (4.7)$$

results in a BCS-like gap. Then, $\Delta' \propto e^{-\lambda/g^2}$, so the phonon process is replaced by pair creation and annihilation through the resonance boson state. As $T_{cdw} \to 0$, both the Boson-induced superfluidity and the resonance boson state significantly diminish allowing BCS superconductivity to prevail.

### Uemura-Homes' law

Uemura's law [10], that the superfluid density $\rho_s$ of the SC cuprates linearly scales to $T_c$ at low doping was one of the first systematic trends in these new materials. Optimally and overdoped materials did not follow this relationship. Later, Homes and others [49] developed a universal scaling relation that holds for a number of high-$T_c$ SC's for the entire SC dome,

$$\rho_s = A\sigma(T_c)T_c$$

where $A$ is a constant and $\sigma(T_c)$ is the normal state conductivity at $T_c$. Using the insightful description of Zaanen [50], we suggest connections of our proposal to Homes' law:

1) Comparing Zaanen's formulas for superconducting and normal states, we have

$\rho_S$:  $\qquad \rho_S \Omega_{p,S}^2 = 4\pi n_S e^2/m_e \qquad\qquad$ superconducting

$\qquad\qquad \Omega_{p,N}^2 = 4\pi n_N e^2/m_e \qquad\qquad$ normal

$\Omega_{p,S}$, $\Omega_{p,N}$ are the plasma frequencies of the SC and normal states, and $n_S, n_N$ are the SC and normal states mobile electron densities, respectively. We suggest the plasma frequencies might relate to the BKT bound vortices (SC) and the CircDW (N) entities which might offer a method of calculating their strengths.

2) $\sigma(T_c) = \Omega_{p,N}^2 \tau(T_c) 8\pi^2$: As Zaanen states, the "normal" state just above $T_c$ might be expected to be viscous can be rationalized by the modified BKT bound vortices which are "melting" over an extended range to $T_D$. Alternatively, Egorov and March's observation [1] supports the increase in viscosity as the LCP are formed just above $T_c$. When the bound vortices completely melt, a lack of balance in the +/- vortices could create another stage of a complex, viscous electronic liquid.

3) The relationship between $\rho_s$ and $T_c$ was discussed earlier.

### 4.4 Comments about the Condensation Energy and a Suggested Gap Equation.

Our first focus in this section is to try and identify the source of the energy of condensation. Marel and others have stated that fluctuations must be counted when the condensation energy is calculated [51]. The model we have presented suggests the existence of several precursor and competing energy states: a LCP, a CircDW, a delocalized CP ($CP_{deloc}$), and possibly a linear CDW at higher energies. This relationship is:

$$LCP \square \ 2circDW\left(^+\!/_-\right) \square \ circDW(+) + circDW(-)$$
$$CP_{deloc}^{2-} \square \ 2e^-$$
$$CircDW \square \ linearDW(CDW)$$

It is not clear to us at this moment how to precisely evaluate the condensation energy. Certainly to make progress it seems apparent that the large entropy associated with the dissociation of a classical BKT bound vortex might offer a starting point [52]. .

### 5. Other Similar Quasi-particles [53].

**5.1 Helium.** Following Landau [54] and Feynman [55], an important property of superfluid helium is that the only low energy excitations possible in superfluid helium are collective density waves, the lack of low-lying single particle excitations occurs because $^4He$ is a boson and assuming we are **below the Landau critical velocity**. Then, the phonons are dissipationless and the low energy excitations of these collective density waves are minimized by a coherent distribution of momentum. If the exact ground state wave function is known, the density wave excited state variational wave function can be written as (at wave vector $\vec{k}$),

$$\Phi_{\vec{k}} = \frac{1}{\sqrt{N}} \rho_{\vec{k}} \psi \qquad (5.1)$$

N being the number of particles and $\rho_{\vec{k}}$, the Fourier transform of the density operator

$$\hat{\rho}_q = \sum_{k,\sigma} c^{\dagger}_{k-q,\sigma} c_{k,\sigma} = \sum_{\vec{r}_i} e^{-i\vec{q}\vec{r}_i} \qquad (5.2)$$

and $\rho_{\vec{k}}$ must have a significant density modulation at $\vec{k}$ resulting in the Bijl-Feynman equation [56]:

$$\Delta(k) = \frac{f(k)}{s(k)} \qquad (5.3)$$

with the norm of the excited state being

$$f(k) = \frac{\langle \psi | \rho_k^{\dagger}(H - E_0)\rho_k | \psi \rangle}{N} \qquad (5.4)$$

($H$ is the Hamiltonian and $E_0$, the ground state). The utility of eq (8) is that the collective mode, a dynamical quantity, is expressed in terms of static properties of the ground state since $f(k)$ is the oscillator strength expressed by the sum rule (Section 2b),

$$f(k) = \frac{\hbar^2 k^2}{2m} \qquad (5.5)$$

Rewriting eq. (5.3) and emphasizing that the Fourier transform of the density-density correlation function can be expressed in the structure factor,

$$S(\vec{k}) = \frac{\langle |\rho_{\vec{k}}|^2 \rangle}{N\rho_0} \qquad (5.6)$$

Landau's roton appears as a damped pole in the density-density correlation function, so the roton will appear as a peak of the structure factor $S(\vec{k})$ vs. the momentum in three dimensions using neutron scattering. Lower dimensions may not have as prominent a scattering, particularly if the numbers of these analogous structures is low (see [57] for an interesting discussion). It appears as though this type of structure is observed experimentally; Greytak et al [58] using Raman scattering found only a single peak near twice the single roton energy. Zawadowski, Ruvalds and Solano (ZRS) [59] then assumed that the roton-roton interaction is attractive and suggested a bound state must exist below the two roton continuum at energy $\varepsilon = 2\Delta_0 - E_b$, $E_b$ being the binding energy, analogous to Cooper pair formation in BCS theory, i.e. a bound state will appear no matter how small the attractive coupling with a binding energy. Greytak later confirmed the binding energy and the d-type angular momentum of the bound state. Another conclusion of this study was the significant impact of the two-particle states on the one-particle spectrum (illustrated in Fig 1a [59]). This feature involves considerable spectral weight transfer, a phenomenon also seen in the cuprates.

**5.2. Fractional Quantum Hall Effect (FQHE).** In the development of effective theories to describe the FQHE Girvin, MacDonald, and Platzman [60] modified Feynman's excitation theory for helium superfluid to describe the collective excitation gap with the analogy that the Hall resistance is almost dissipation-less. It proved especially interesting when Rezayi and Haldane found quantitative evidence for off-diagonal long range order (ODLRO) in Quantum Hall States [61].

**5.3. BEC.** In cold-atom BEC studies the roton arises through anisotropic dipole-dipole interactions (density-density correlation) which are partially attractive and result in two rotons binding in with a "d" symmetry. These features are absent in repulsive short-ranged s-wave interactions. O'Dell and others [62] have been able to drive the transition by laser intensity as illustrated by the BEC structure factor $S(k)$ as further discussed by Minguzzi, March and Tosi [63]. This type of structure is particularly interesting to study as the roton minimum has been suggested to being very close to incipient crystallization as the ordering of atoms leads to a peak in $S(k)$ and perhaps, ultimately, a solid-fluid transition, as discussed by Nozieres [64].

**5.4. Comparison of KT and other similar quasi-particles.**

There are other comparisons between these quasi-particles and the LCP-BKT model of superconductivity (see [53] for those not discussed here). It is plausible that the BKT Cooper pair has a "d" symmetry, similar to the roton-roton interactions discussed above. Secondly, and more specifically, Magro and Ceperley [65] have noted that the 2D Bose Coulomb liquid (2DBCL) contains long range correlations similar to the bosonic representation of Laughlin's wave function and identical to Girvin and MacDonald's algebraic long-range order. It is a different matter at short range as Laughlin's wave function for two particles is classical in nature while quantum mechanical 2DBCL results in a depressed but finite probability for this comparison. Both Magro et al and Minguzzi et al recognize that superfluidity can result even if there is no BEC.

Other systems of interest include Fogler and Koulakov [66] who have compared the energy of the Laughlin liquid and a charge density wave in a weak magnetic field. They conclude that after optimization of the CDW that it is lower in energy for Landau level numbers $N \geq 2$ at $v_N = \frac{1}{3}$ which agrees with experiment. Cox and others [67] have noted that a CDW coexisting with disorder may be important in manganite superstructures. Certainly the topological nature of QHE and the LCP-BKT bear closer scrutiny.

**6. Summary, Conclusions, and Proposals for Future Work.**

**6.1 Summary.** Based on earlier work, we have proposed a microscopic model for cuprate HTSC by relaxing the "energy shell" in Cooper's model ($\Box \, \varepsilon_F \pm \hbar \omega_D$) [see Appendix]. This allows us to pursue a real space Cooper pair which we identify as the "preformed pair." We suggest that these pairs form a BKT bound vortex which dissociates above $T_c$ into circular CDW's. All of these structures are highly correlated. As a function of temperature, doping and magnetic field strength, there seem to be at least five distinct zones in the phase diagram. The model also offers an explanation of several disparate experimental observations – Nernst effect, bosons above $T_c$, electrons and vortices interacting, etc.

**6.2 Conclusions.** The model seems to have the ability to rationalize a number of widely differing experimental observations from STM and Nernst effects to "local superconductivity persisting to zero temperature" [8]. It also seems to have the ability to possibly explain several experimental and theoretical criteria that heretofore seemed beyond reach such as the Uemura-Homes law and the density postulates of Turlakov and others. Certainly a calculation of condensation energy would be helpful and with this model perhaps is can be calculated. It is intriguing that doping a LCP system, being a BKT vortex, is therefore a topological object; doping it leads to SC. Is this a general statement?

While the title refers solely to high $T_c$ cuprates, there are relevant but limited comparisons with (a) heavy-Fermion superconductors and (b) fullerides. The parallels of heavy-Fermion superconductor $CeCoIn_5$ with high $T_c$ cuprates are numerous: (i) the quasi-two-dimensionality; (ii) the d-wave superconductivity; (iii) the appearance of superconductivity in the neighborhood of an antiferromagnetic state [68]. In the study of Dora et. al. the large Nernst effect observed by R. Bel [69] above $T_c \square 2.3K$ in $CeCoIn_5$ is interpreted in terms of an unconventional density wave (UDW) which is in evidence around $T = 18K$. Dora et al also offer an explanation of the temperature dependence of the Seebeck coefficient below $18K$ in terms of a UDW.

**Proposals for Future Work.**

1. The structures proposed can be readily solved numerically. It would be quite interesting to insert some of these structures into one of the several sophisticated programs to see if quantitative results can be obtained.

2. It would be interesting to explore similar structure that appear in other materials which are not SC's; why?

**Appendix. "Soft" Cooper Pair Boson.**

Our initial notion of a CP was fixed by BCS theory approximations. A CP emerging from the continuum initially seemed to have a bit of mystery about its origin but bound states emerging from the continuum have appeared elsewhere [59]) and there is a sound basis [53]. The matrix element which leads to the attraction which makes the Fermi sea unstable is well known, namely, eq (A.1)

$$V(\vec{k}, \vec{k} - \vec{q}) = \frac{\hbar \varpi}{\left\{\varepsilon(\vec{k}) - \varepsilon(\vec{k} - \vec{q})\right\}^2 - (\hbar \varpi)^2} \qquad (A.1)$$

approximated by BCS as $V_{k,k'} = -V < 0$ if $|\varepsilon_k|, |\varepsilon_{k'}| < \omega_c$ and zero otherwise (a particle in box potential). With this premise of an attraction of strength V, no matter how weak, Cooper diagonalized a very large matrix representing an m-body problem results in M-1

$$\hat{H}_{pair} = \begin{pmatrix} 2\varepsilon & -V & -V & . \\ -V & 2\varepsilon & -V & . \\ -V & -V & 2\varepsilon & . \\ . & . & . & . \end{pmatrix} \quad \text{using } \hat{H}^T = \hat{S}^{-1}\hat{H}_{pair}S \text{ with } S_{jk} = \exp(2\pi ijk/n). \quad (A.2)$$

degenerate levels being raised in energy and a single non-degenerate level, the Cooper pair, lower in energy,

$$\hat{H}^T = \begin{pmatrix} 2\varepsilon - nV & 0 & 0 & . \\ 0 & 2\varepsilon & 0 & . \\ 0 & 0 & 2\varepsilon & . \\ . & . & . & . \end{pmatrix} \quad (A.3)$$

All M electron pairs of the same momentum can scatter and obey the exclusion principle, giving maximum correlation.

Removing the energy requirement of the BCS approximation, i.e. letting $\Delta\varepsilon = \varepsilon(\vec{k}) - \varepsilon(\vec{k}-\vec{q}) - \hbar\varpi$ be small, the perturbation term in eq (A.1) breaks down, which provides the condition for real (as opposed to virtual) electron-phonon interactions. The "energy shell" that results can be much smaller than $\hbar\omega_D$. Eq (A.1) can still be used by taking the limit $\Delta\varepsilon \to 0$ by interpreting the pole as a principle value [70]. Applying the same concept to a circCDW and using a proof of the Bloch theorem or spin system a four-membered ring with the distance between members, a, satisfies a wave equation $\psi(x+Na) = \psi(x) = C^N\psi(x)$. Assuming the interaction V is only between nearest neighbors the interaction matrix is

$$\hat{H}_{circ} = \begin{pmatrix} \varepsilon & \beta & 0 & \beta \\ \beta & \varepsilon & \beta & 0 \\ 0 & \beta & \varepsilon & \beta \\ \beta & 0 & \beta & \varepsilon \end{pmatrix} \quad (A.4)$$

The secular equation can be solved by $\beta c_{r-1} - \varepsilon c_r + \beta c_{r+1} = 0$. A single-valued $\psi$ requires C to be one of the N roots of unity

$$C = \exp(i2\pi j/N) \text{ with } j = 0, \pm 1, ...$$

This solution always has a paired lowest energy state with a dimensionless Fröhlich coupling constant, $\lambda$. Richardson has suggested another approach [71].